\documentclass[journal,twoside,web]{ieeecolor}
\usepackage{generic}
\usepackage{cite}
\usepackage{amsmath,amssymb,amsfonts}
\usepackage{algorithmic}
\usepackage{graphicx}
\usepackage{algorithm,algorithmic}
\usepackage{hyperref}
\usepackage{booktabs}
\hypersetup{hidelinks=true}
\usepackage{textcomp}
\def\BibTeX{{\rm B\kern-.05em{\sc i\kern-.025em b}\kern-.08em
    T\kern-.1667em\lower.7ex\hbox{E}\kern-.125emX}}
\begin{document}
\title{DiagR1: A Vision-Language Model Trained via Reinforcement Learning for Digestive Pathology Diagnosis}

\author{Minxi Ouyang, Lianghui Zhu, Yaqing Bao, Qiang Huang, Jingli Ouyang, Tian Guan, Xitong Ling, Jiawen Li, Song Duan, Wenbin Dai, Li Zheng, Xuemei Zhang, Yonghong He
\thanks{Manuscript received xx xx, xxxx; revised xx xx, xxxx.. This study was supported by grants from the National Natural Science Foundation of China (82160108)}
\thanks{Minxi Ouyang, Lianghui Zhu, Jingli Ouyang, Xitong Ling, Jiawen Li, Tian Guan, Yonghong He are with Shenzhen International Graduate School, Tsinghua University, Beijing, China.}
\thanks{Xuemei Zhang and Wenbin Dai are with Department of Pathology, Liuzhou People's Hospital Affiliated to Guangxi Medical University, Liuzhou, Guangxi, China.}  
\thanks{Li Zheng is with Department of Immunology, College of Basic Medical Sciences, China Medical University, Shenyang, Liaoning Province, P.R. China.}
\thanks{Yaqing Bao is with Greater Bay Area Center for Medical Device Evaluation and Inspection.NMPA, Shenzhen, Guangdong Province,P.R. China.}
\thanks{Qiang Huang is with Shenzhen Shengqiang Technology Co., Ltd., Shenzhen, Guangdong, China.}
\thanks{Song Duan is with Department of Pathology, Chongqing University Affiliated Three Gorges Hospital, Chongqing, China.} 
\thanks{Minxi Ouyang, Lianghui Zhu, Yaqing Bao contributed equally to this work.}
\thanks{Corresponding author: Yonghong He (e-mail: heyh@sz.tsinghua.edu.cn), Xuemei Zhang (e-mail: zhangxuem202109@163.com), Li Zheng
 (e-mail: lzheng20@cmu.edu.cn)}
}

\maketitle

\begin{abstract}
Multimodal large models have shown great potential in automating pathology image analysis. 
However, current multimodal models for gastrointestinal pathology are constrained by both data quality and reasoning transparency: pervasive noise and incomplete annotations in public datasets predispose vision–language models to factual hallucinations when generating diagnostic text, while the absence of explicit intermediate reasoning chains renders the outputs difficult to audit and thus less trustworthy in clinical practice.
To address these issues, we construct a large-scale gastrointestinal pathology dataset containing both microscopic descriptions and diagnostic conclusions, and propose a prompt argumentation strategy that incorporates lesion classification and anatomical site information. This design guides the model to better capture image-specific features and maintain semantic consistency in generation. Furthermore, we employ a post-training pipeline that combines supervised fine-tuning with Group Relative Policy Optimization (GRPO) to improve reasoning quality and output structure. Experimental results on real-world pathology report generation tasks demonstrate that our approach significantly outperforms state-of-the-art open-source and proprietary baselines in terms of generation quality, structural completeness, and clinical relevance. Our solution outperforms state-of-the-art models with 18.7\% higher clinical relevance, 32.4\% improved structural completeness, and 41.2\% fewer diagnostic errors, demonstrating superior accuracy and clinical utility compared to existing solutions.
\end{abstract}


\section{Introduction}
\label{sec:introduction}
\IEEEPARstart{W}{ITH} the clinical approval and large-scale deployment of digital whole-slide image (WSI) systems worldwide, traditional field-by-field microscopic examination is gradually being replaced by computational “pixel-level pathology.” This transformation has laid the foundation for the application of deep learning models in histological diagnosis by enabling high-resolution data acquisition and standardized workflows \cite{huang2025computational}\cite{2}. The burden of gastrointestinal (GI) diseases continues to rise: recent epidemiological studies show that, in 2024 alone, the number of medical visits and related examinations due to GI conditions in the United States has increased significantly, accompanied by a surge in biopsy specimen volume\cite{3}\cite{zena2024patterns}\cite{yadlapati2022quality}. In stark contrast, the number of qualified pathologists has grown slowly and remains unevenly distributed across the globe \cite{walsh2024current}. Against this backdrop of supply–demand imbalance, there is an urgent need to develop computational pathology systems capable of automatically detecting lesions and generating structured diagnostic reports \cite{liang2024enhancing}\cite{ren2025application}.


In recent years, general vision-language large models(VLLMs) (such as Qwen-VL\cite{bai2025qwen2}, DeepSeek-VL\cite{guo2025deepseek}, etc.) have sparked a new wave of technological innovation in the field of pathology, leveraging their ability to simultaneously understand and generate both visual and linguistic information. Compared to traditional methods that rely solely on convolutional networks, these models can perform diagnostic-level visual analysis and structured text generation within a single framework, providing an end-to-end solution for pathology workflows, from "viewing images" to "writing reports"\cite{ye2021medpath}\cite{tran2025generating}. Despite these advancements, we identify two key bottlenecks that hinder the practical deployment of computational pathology in gastrointestinal applications\cite{dos2022computational}\cite{rau2024closing}:

\begin{figure*}[!t]
\centering
\includegraphics[width=\textwidth]{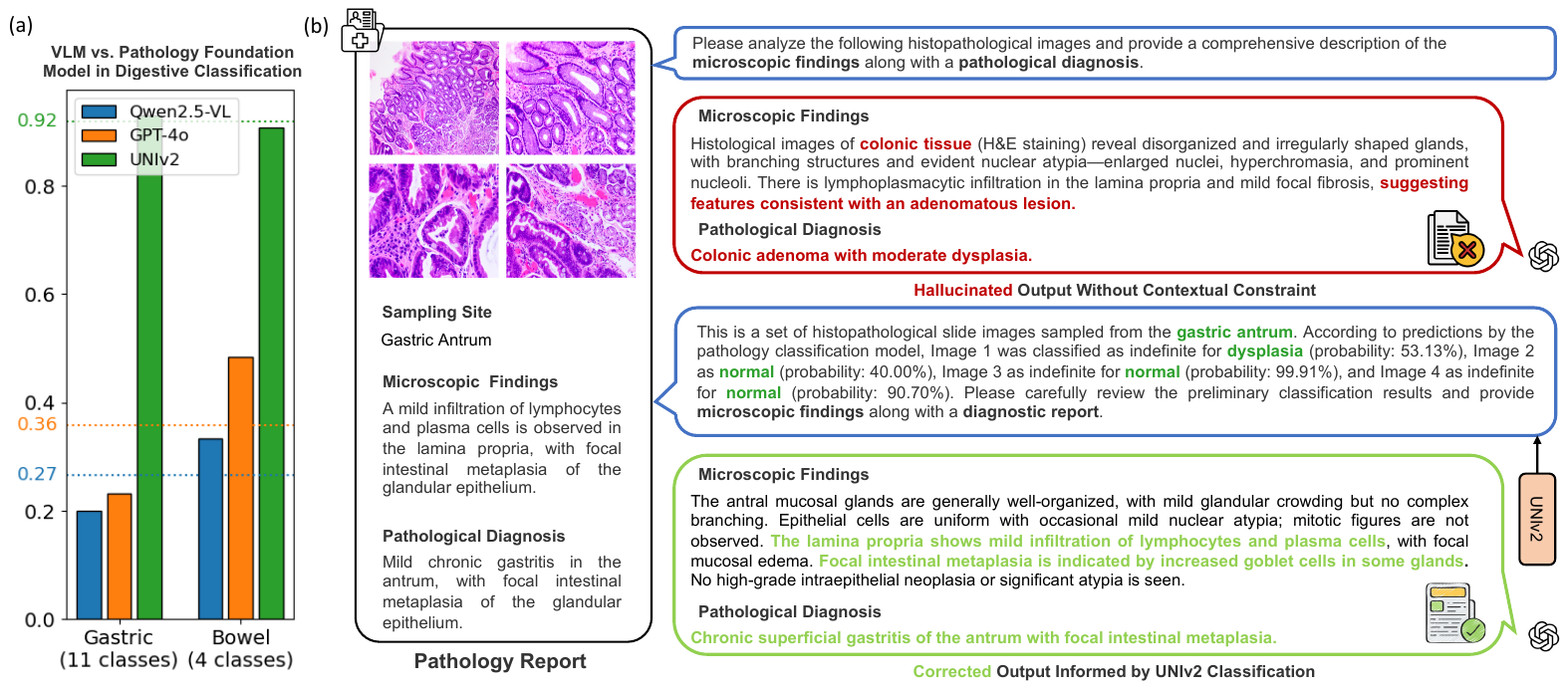}
\caption{
Challenges of general VLLMs in pathological image analysis. (a) Performance comparison between VLLMs and the pathology foundation model in digestive classification tasks.
(b) Impact of classification prompts from the foundation model on pathology report generation.
}
\label{start}
\end{figure*}

\begin{enumerate}

    \item Due to the substantial noise and incomplete annotations present in existing datasets, current vision-language models (VLMs) often generate factual inconsistencies, commonly referred to as "hallucinations" \cite{rajaganapathy2025synoptic}\cite{asgari2024framework}. As shown in Fig.~\ref{start}, in our custom-designed 11-class gastric and 4-class intestinal classification tasks, pathology-specific visual models exhibit significantly higher classification accuracy compared to general-purpose VLMs. Moreover, once a general VLM makes an incorrect interpretation of a pathological image, it tends to continue generating text along that erroneous line of reasoning, exacerbating the risk of hallucinated content. 


    \item Current multimodal generative models used for pathology analysis often lack intermediate reasoning evidence, which results in outputs that lack sufficient traceability, thereby reducing clinicians' trust. Existing public datasets and clinical pathology reports frequently omit critical reasoning steps, posing a significant challenge for effective model training. At the same time, to ensure the completeness of key information, surgical and gastrointestinal pathology departments commonly adopt structured templates for report writing. However, this practice leads to highly homogeneous language patterns, making language models prone to "template-style overfitting"\cite{kefeli2023benchmark}\cite{schaad2024impact}\cite{eloy20241}. 
\end{enumerate}

To address the issues mentioned above, we collected and curated 18,627 gastrointestinal pathology reports, among which 11,904 reports contain complete microscopic descriptions, while the remaining 6,723 include only diagnostic conclusions. Building on recent advances in reasoning-enhanced training demonstrated by models such as DeepSeek-R1~\cite{guo2025deepseek}, we first performed supervised fine-tuning (SFT) on these pathology reports to warm-start the model, followed by reinforcement learning using the GRPO algorithm. In addition, considering the high classification accuracy of pathology-specific foundation models at the ROI level, we designed a prompt argumentation module: a high-performance classifier is used to automatically infer the lesion category from the WSI, which, along with biopsy site information, is embedded into the prompt to guide the language model to focus on key morphological features when generating “microscopic findings.” Incorporating these two components, we propose a new model, \textbf{DiagR1}.

We evaluated our proposed framework on real-world pathology report generation tasks, including both “microscopic findings” and “pathological diagnostic” sections. Experimental results demonstrate that our framework achieves substantial improvements on both internal and external test sets. Compared to the baseline model GPT-4o, it achieves a 25.55\% performance gain, indicating its strong potential for deployment in clinical gastrointestinal pathology report generation.

\begin{enumerate}

    \item We utilize reinforcement learning (RL) to enhance the generation of gastrointestinal pathology reports using multimodal large models. This approach leverages diagnostic information to implicitly improve the quality and richness of the "microscopic findings" generation, thereby enhancing the clinical interpretability of the reports.
    
    \item We propose a new prompt argumentation strategy that utilizes high-accuracy pathology foundation models to provide structured guidance for multimodal large models, enabling better comprehension of image context and mitigating the adverse effects of hallucination.
    
    \item We conduct both internal and external validations of our model, demonstrating significant improvements in the accuracy of generating both microscopic findings and diagnostic conclusions.

\end{enumerate}

\begin{figure*}[!t]
\centering
\includegraphics[width=\textwidth]{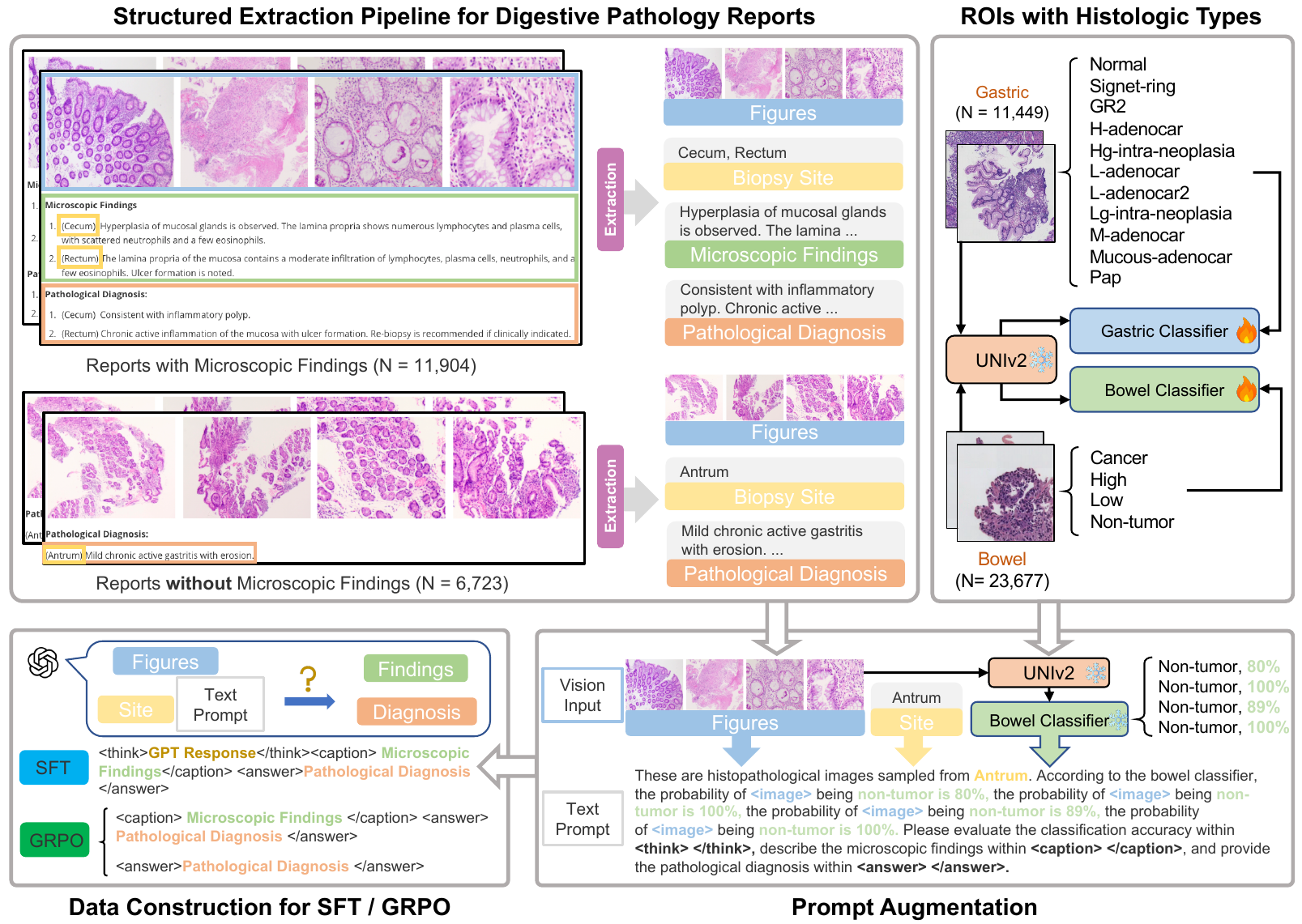}
\caption{
Pipeline for constructing a digestive pathology report dataset. The process involves two main stages: structured extraction of key information from raw pathology reports and the construction of an ROI classification dataset to train pathology classifiers. The trained classifiers are then used to predict on image-text samples, and the classification results are incorporated into the original text for prompt augmentation, generating reasoning-chain data for subsequent SFT or GRPO training.
}
\label{data}
\end{figure*}

\section{Related Works}

\subsection{Advancements in Visual Encoders for Pathology}
In the field of computational pathology, optimizing visual encoders to enhance pathological models' understanding of histopathological images has emerged as a research priority \cite{ding2024multimodal}\cite{vorontsov2024foundation}. The UNI\cite{chen2024towards} model, based on the Transformer framework, employs contrastive learning or masked modeling strategies to improve generalized feature representation of cellular morphology and tissue structures. CTransPath\cite{wang2022transformer} utilizes dual-branch attention mechanisms or feature modulation modules to model relationships between cellular-level details and tissue-level context, thereby enhancing multi-scale semantic comprehension efficiency. CONCH\cite{lu2024visual}, trained on 1.17 million image-text pairs through contrastive learning, achieves state-of-the-art performance in zero-shot classification and cross-modal retrieval across 14 benchmark tasks, surpassing specialized models. PLIP\cite{huang2023visual}, fine-tuned from the CLIP \cite{radford2021learning} architecture using 200,000 pathology image-text pairs, demonstrates exceptional zero-shot capabilities in rare disease classification. These advancements collectively enhance AI models' pathological image comprehension, driving the evolution from single-task solutions to versatile foundation models.

\subsection{Large Vision Language Models in Pathology}

The rapid development of multimodal large language models (MLLMs)\cite{liu2023visual}\cite{liu2024improved}\cite{bai2025qwen2} has facilitated the integration of large language models (LLMs) with computational pathology foundation models, yielding numerous pathological MLLMs\cite{sun2024pathasst}\cite{ma2024towards}\cite{xu2024multimodal}\cite{luo2022biogpt}. Optimizing decoder architectures has become a key strategy for improving pathological report generation and cross-modal alignment. PathChat\cite{lu2024multimodal}, a specialized multimodal assistant for pathology, enhances diagnostic accuracy beyond general models like GPT-4V\cite{openai20234v} through refined decoder-based multimodal fusion modules. Quilt-LLaVA\cite{seyfioglu2024quilt} constructs instruction-tuning datasets from localized narratives of open-source pathology videos, achieving superior performance in cross-modal retrieval and question-answering tasks compared to traditional supervised methods. HistoGen\cite{guo2024histgen} integrates local-global hierarchical encoders with cross-modal context modules. 

\subsection{Reinforcement Learning for Post-training VLMs}

 Recently, reinforcement learning \cite{shen2025vlm}\cite{huang2025vision} has emerged as a key paradigm for enhancing the reasoning capabilities of VLMs\cite{guo2025deepseek}\cite{jaech2024openai}, with its applications expanding from language models (e.g., OpenAI's o1) to multimodal domains. LLaVA-CoT\cite{thawakar2025llamav} employs a Chain-of-Thought (CoT) enhancement strategy, leveraging structured multimodal reasoning datasets for instruction tuning to improve traceability in reasoning processes; Visual-RFT\cite{liu2025visual} focuses on direct perception optimization by designing verifiable visual reward functions to enhance generalization under limited supervision in fine-grained classification tasks for natural images. Reinforcement learning (RL) has demonstrated promising applications in the medical domain\cite{liu2024histogym}\cite{raza2024dual}, such as enhancing the cross-task generalization capabilities of VLMs across eight medical imaging modalities through GRPO\cite{lai2025med}. However, more systematic modeling approaches remain imperative for specialized scenarios like pathology analysis.

\begin{figure*}[th]
\centering
\includegraphics[width=\textwidth]{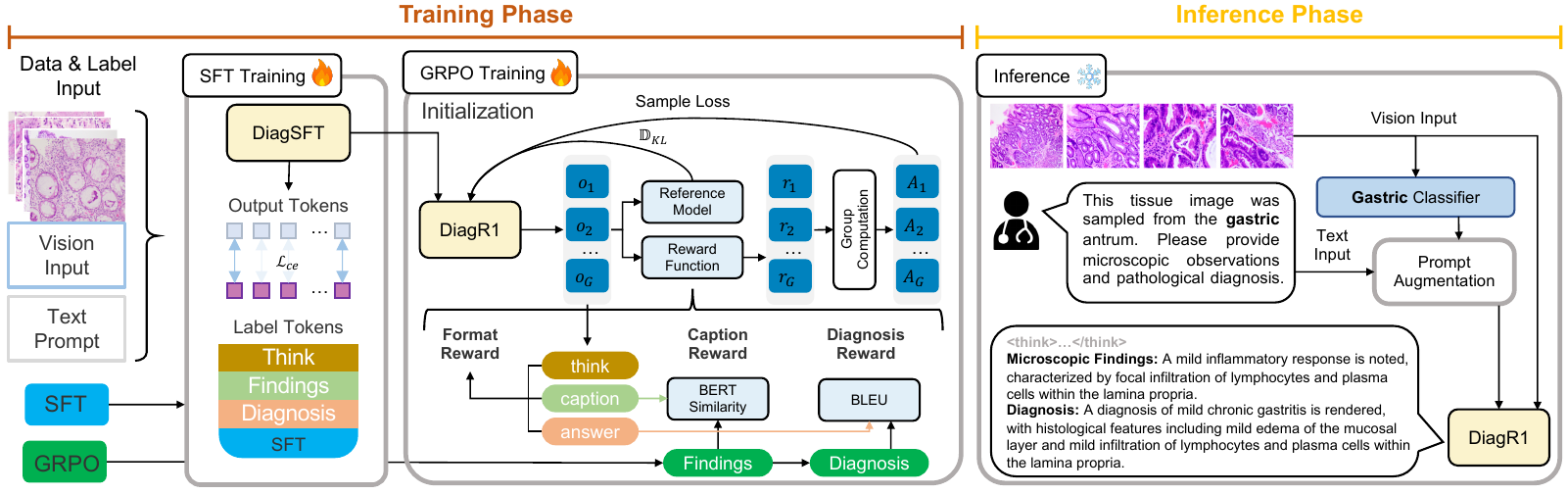}
\caption{
Overview of the proposed DiagR1 training and inference framework. In the training phase, enhanced image-text data are used for SFT and reinforcement learning based on GRPO, where task-specific reward functions are designed to optimize the quality of generated microscopic findings and diagnoses for digestive pathology. This training pipeline enables DiagR1 to generate pathology reports efficiently and reliably during inference.
}
\label{pipeline}
\end{figure*}

\section{Methodology}

\subsection{Dataset Construction}

To improve the performance of multimodal diagnostic models in digestive pathology report generation, this study constructs a specialized dataset for the digestive system from two perspectives: structured clinical semantics and clinically relevant classification annotations. The dataset leverages authentic and diverse pathology reports for model training and validation, while incorporating clinically adopted classification standards to annotate ROIs, thereby enhancing the model’s generalization capability. The detailed pipeline, including the structured extraction pipeline for digestive pathology reports and ROIs with histologic types, is illustrated in Fig.~\ref{data}.

\textbf{Structured Extraction Pipeline for Digestive Pathology Reports.} We collected gastrointestinal pathology reports from two Class-A tertiary hospitals, Liuzhou People's Hospital Affiliated to Guangxi Medical University and Chongqing University Affiliated Three Gorges Hospital. Liuzhou People's Hospital Affiliated to Guangxi Medical University secured ethical approval and signed data-use agreements(No. KY-2025-232). All slides and reports were fully de-identified before acquisition and anonymized with random identifiers. After extracting ROI images, sampling sites, microscopic findings, and diagnostic conclusions, we obtained 11,904 reports with complete microscopic descriptions (47,258 images) and 6,723 reports containing only diagnostic conclusions (23,532 images); notably, all samples from Chongqing University Affiliated Three Gorges Hospital lacked microscopic findings.

\textbf{ROIs with Histologic Types.} We curated 11,449 representative gastric ROIs and 23,677 bowel ROIs, each independently annotated by board-certified gastrointestinal pathologists. These ROIs were rigorously curated from four multicenter medical institutions and encompassed 11 distinct histopathological diagnostic categories. Class 0 included non-neoplastic lesions such as gastritis, intestinal metaplasia, reactive hyperplasia, fundic gland polyps, and hyperplastic polyps (432 ROIs). Class 1 and Class 2 corresponded to LIN (1,139 ROIs) and HIN (722 ROIs), respectively. Classes 3 to 9 included various gastric carcinomas: well-differentiated adenocarcinoma (1,273 ROIs), moderately differentiated adenocarcinoma (2,042 ROIs), poorly differentiated adenocarcinoma (1,749 ROIs), signet-ring cell carcinoma (1,085 ROIs), mucinous carcinoma (796 ROIs), other poorly cohesive carcinomas (756 ROIs), papillary adenocarcinoma (837 ROIs) and atypical hyperplasia (618 ROIs). The intestinal images were grouped into 4 classes: carcinoma (5,165 ROIs), high-grade intraepithelial neoplasia (844 ROIs), low-grade intraepithelial neoplasia (6,459 ROIs), and non-neoplastic tissue (11,209 ROIs). 

\subsection{Prompt Augmentation}

To reduce the risk of hallucinations in general-purpose multimodal large models applied to digestive pathology tasks, we introduce auxiliary prompts derived from the classification capabilities of a domain-specific foundation model. Specifically, regions of interest (ROIs) from gastric and intestinal tissue samples are collected and uniformly resized to 512 pixels along the longer edge. Visual features are extracted using a frozen UNIv2 model, yielding 1,536-dimensional representations. Separate linear classification heads with dimensions $(D, C)$, where $D=1536$ and $C$ denotes the number of classes for each organ, are trained to construct a dual-branch ROI classifier (Fig.~\ref{data}). During inference, the system selects the appropriate classifier branch based on the anatomical site and performs rapid ROI classification, retrieving the top-ranked category along with its associated confidence score. This predicted label and confidence, together with the site of origin, form an auxiliary prompt that is concatenated with the structured prompt before being fed into the multimodal model. For samples from organs beyond the stomach or intestine, only the anatomical site is used to enhance source-awareness in the prompt design.

\subsection{RL Post-training for Digestive Pathology Reports}

\subsubsection{Warm-start Phase with Supervised Fine-tuning}

Before initiating the reinforcement learning phase based on GRPO, we employ a warm-start strategy via SFT to initialize the multimodal pathology model. This design is motivated by three main considerations. First, directly exploring the complex task of pathology report generation using a random policy often leads to a “cold start” problem, characterized by sparse rewards, unstable value estimation, and non-convergent training. Prior work has shown that incorporating offline demonstrations or behavior cloning significantly improves sample efficiency and stabilizes learning dynamics~\cite{guo2025deepseek}. Second, in the training of large language models, SFT has been widely adopted as the standard initialization step before reinforcement learning. Most RLHF (Reinforcement Learning from Human Feedback) paradigms incorporate an SFT phase to inject prior knowledge, enhance sample efficiency, and ensure stable policy optimization~\cite{thawakar2025llamav}~\cite{liu2025visual}. Third, to guarantee that subsequent reward signals are interpretable across the reasoning chain, we adopt an explicit chain-of-thought prompting scheme, where \texttt{<think>...</think>} encodes intermediate reasoning, \texttt{<caption>...</caption>} captures microscopic observations, and \texttt{<answer>...</answer>} provides the final diagnostic output. The pipeline is illustrated in Fig.~\ref{data}.

Motivated by these considerations, we construct a two-stage data generation pipeline for SFT: (i) The image, prompt argumentation structured text, original microscopic findings, and diagnostic conclusions are fed into a state-of-the-art multimodal model to guide the generation of reasoning traces in the \texttt{<think>} section, focusing on ROI-level morphological and semantic logic; (ii) The \texttt{<caption>} and \texttt{<answer>} sections are then populated with expert-verified descriptions and diagnoses, resulting in structurally complete and hierarchically labeled SFT training instances.

\subsubsection{Diagnosis-guided RL Post-training}

During the data construction phase for reinforcement learning, we selected a set of samples that are entirely disjoint from those used in the SFT stage. The dataset includes both cases with available microscopic findings and those lacking such descriptions, thereby ensuring diversity in the training data. Meanwhile, we adopt the same data construction protocol as in the SFT phase, using the original pathology reports as supervision labels for training.

GRPO is a variant of Proximal Policy Optimization (PPO)~\cite{schulman2017proximal}  that replaces the traditional reward model with a group-based baseline mechanism, significantly reducing computational cost during training. Specifically, for each input query $q$, GRPO samples a group of outputs $\{o_1, o_2, \cdots, o_G\}$ from the old policy $\pi_{\theta_{\text{old}}}$, and optimizes the current policy $\pi_\theta$ by maximizing the following objective:

\begin{equation}
\begin{aligned}
J_{\text{GRPO}}(\theta) 
&= \mathbb{E}_{q \sim P(Q),\ \{o_i\}_{i=1}^G \sim \pi_{\theta_{\text{old}}}(O|q)} \\
& \Bigg[ \frac{1}{G} \sum_{i=1}^G \min\Bigg( \frac{\pi_\theta(o_i \mid q)}{\pi_{\theta_{\text{old}}}(o_i \mid q)} A_i, \\
&\quad \text{clip}\left( \frac{\pi_\theta(o_i \mid q)}{\pi_{\theta_{\text{old}}}(o_i \mid q)},\ 1 - \varepsilon,\ 1 + \varepsilon \right) A_i \Bigg) \\
&\quad - \beta D_{\text{KL}}(\pi_\theta \parallel \pi_{\text{ref}}) \Bigg]
\end{aligned}
\end{equation}

The KL divergence term in Equation~(1) is defined as:
\begin{equation}
\mathbb{D}_{\text{KL}}(\pi_\theta \,\|\, \pi_{\text{ref}}) = 
\frac{\pi_{\text{ref}}(o_i \mid q)}{\pi_\theta(o_i \mid q)} 
- \log \frac{\pi_{\text{ref}}(o_i \mid q)}{\pi_\theta(o_i \mid q)} - 1
\tag{2}
\end{equation}

Here, $\varepsilon$ and $\beta$ are hyperparameters, and $A_i$ denotes the advantage term, which is computed by standardizing the reward values $\{r_1, r_2, \dots, r_G\}$ within the sampled output group:
\begin{equation}
A_i = \frac{r_i - \text{mean}(\{r_1, r_2, \cdots, r_G\})}{\text{std}(\{r_1, r_2, \cdots, r_G\})}
\tag{3}
\end{equation}

\subsubsection{Reward Function for Pathology Report Generation}

In reinforcement learning, the design of reward functions plays a critical role in guiding model behavior and directly impacts the quality and task alignment of generated outputs. Well-defined rewards not only improve task-specific performance but also enforce desirable structural and semantic properties in the generation process. To meet the unique requirements of digestive pathology report generation, we design a composite reward framework comprising three components: a format reward to enforce structural consistency, a caption reward to enhance microscopic reasoning, and a diagnosis reward to improve diagnostic accuracy.

\textbf{Format Reward.}  
The format reward $R_{\text{format}}$ is designed to ensure that the generated report strictly adheres to a predefined structure. The output must include three segments: reasoning enclosed within \texttt{<think>...</think>}, microscopic findings within \texttt{<caption>...</caption>}, and the final diagnostic conclusion within \texttt{<answer>...</answer>}. A reward of 1 is assigned only if all three tags appear exactly once and all content is properly enclosed within these tags without extraneous text. If any structural constraint is violated, such as missing, duplicated, or misaligned tags, the reward is set to 0.

\begin{table*}[th]
\centering
\caption{Evaluation results of various general models and our model DiagSFT and DiagR1 with prompt augmentation on the LZ and CQ gastrointestinal pathology datasets. The best-performing methods are \textbf{bolded}}
\label{tab:evaluation1}
\begin{tabular}{lcccccc}
\toprule
\textbf{Model} & \multicolumn{2}{c}{\textbf{Microscopic Findings (LZ)}} & \multicolumn{2}{c}{\textbf{Diagnosis (LZ)}} & \multicolumn{2}{c}{\textbf{Diagnosis (CQ)}} \\
\cmidrule(lr){2-3} \cmidrule(lr){4-5} \cmidrule(lr){6-7}
& BERT Score (\%) & BLEU (\%) & BERT Score (\%) & BLEU (\%) & BERT Score (\%) & BLEU (\%) \\
\midrule
GPT-4o & 82.66 & 1.12 & 85.07 & 2.69 & 84.37 & 2.39 \\
Gemini-2.5-pro & 79.21 & 0.71 & 78.36 & 0.79 & 78.30 & 1.83 \\
Llama-3.2-11B-Vision & 80.23 & 0.09 & 80.00 & 0.10 & 80.31 & 0.21 \\
DeepSeek-VL2 & 78.31 & 0.05 & 77.65 & 0.08 & 77.21 & 0.10 \\
Qwen2.5-VL-7B & 81.35 & 1.15 & 79.65 & 0.90 & 81.13 & 1.53 \\
\midrule
DiagSFT & 86.38 & 7.55 & 88.39 & 7.61 & 84.73 & 3.42 \\
DiagR1 & \textbf{88.55} & \textbf{12.95} & \textbf{92.80} & \textbf{20.93} & \textbf{86.65} & \textbf{4.20} \\
\bottomrule
\end{tabular}
\end{table*}

\textbf{Caption Reward.}  
Once the output structure is validated, the model generates the microscopic findings in the \texttt{<caption>} segment. To encourage informative and semantically faithful content, we adopt a similarity-based reward using a domain-specific BERT encoder. Let $c_i$ denote the generated caption extracted from output $o_i$, and $c$ be the reference caption. We compute their cosine similarity based on their BERT embeddings $\mathbf{e}_{c_i}$ and $\mathbf{e}_c$:
\begin{equation}
R_{\text{cap}}(c_i, c) = \frac{\mathbf{e}_{c_i} \cdot \mathbf{e}_c}{\|\mathbf{e}_{c_i}\| \, \|\mathbf{e}_c\|}
\tag{5}
\end{equation}
This similarity score serves as a soft reward signal that encourages semantic alignment between generated and reference findings, while also promoting diversity in expression.

\textbf{Diagnosis Reward.}  
To ensure the generated diagnosis is accurate and linguistically coherent, we adopt the BLEU score as the reward for the \texttt{<answer>} segment. Let $r$ be the reference diagnosis and $d_i$ the model-generated diagnosis from $o_i$. The BLEU score is computed by averaging the log-precision of 1-gram to 4-gram matches with equal weights $w_n$:
\begin{equation}
R_{\text{ans}}(r, d_i) = \exp \left( \sum_{n=1}^{4} w_n \cdot \log p_n(r, d_i) \right)
\tag{6}
\end{equation}
Here, $p_n(r, d_i)$ denotes the $n$-gram precision between the reference and hypothesis. Optimizing this reward encourages the model to produce diagnostic statements that are both fluent and consistent with ground truth reports.

The final reward is computed as a weighted sum of the three components:
\begin{equation}
r_i = \lambda_{\text{format}} R_{\text{format}}(o_i) 
+ \lambda_{\text{cap}} R_{\text{cap}}(c_i, c) 
+ \lambda_{\text{ans}} R_{\text{ans}}(r, h_i)
\tag{7}
\end{equation}
where $\lambda_{\text{format}}$, $\lambda_{\text{cap}}$, and $\lambda_{\text{ans}}$ 
are non-negative weighting coefficients that control the relative importance 
of the format, caption, and diagnosis rewards, respectively.

\section{Experiments}

\subsection{Dataset Split and Evaluation Strategy}

For the ROI classification task on gastric and intestinal datasets, we follow the official split strategy of UNIv2, dividing all samples into a training set and a test set with an 8:2 ratio. Only the training set is used to fine-tune the linear classification head, while performance is evaluated on the test set to ensure no data leakage and maintain consistency with prior work.

For the pathology report data, we consider the differences in the completeness and formatting of the "microscopic findings" field across two hospitals. Due to the limited size and missing caption information in samples from Chongqing University Affiliated Three Gorges Hospital, all such data are allocated to the test dataset to promote cross-institutional generalization, defined as CQ. For data from Liuzhou People's Hospital, Affiliated to Guangxi Medical University, we stratify the samples as follows: 3,000 reports with complete microscopic findings are selected as supervision data for the SFT stage; another 3,000 similar reports are used as an independent test dataset for evaluating both diagnosis and caption generation, defined as LZ; the remaining 12,627 reports, regardless of caption availability, are used in GRPO training to enrich policy exploration diversity.

For evaluation, we adopt a dual-perspective protocol: following HistoGPT\cite{tran2025generating}, we first encode model outputs using a pre-trained BERT model and compute cosine similarity to assess semantic alignment. We then calculate BLEU-1 to BLEU-4 scores (with uniform n-gram weights of 0.25) to evaluate surface-level fidelity, jointly measuring the quality of generated reports.

\subsection{Implementation Details}

We adopt Qwen2.5VL-7B-Instruct \cite{bai2025qwen2} as our base model, a high-performance, lightweight open-source multimodal language model. In both the SFT and GRPO stages, we perform full-parameter fine-tuning. During the SFT phase, training is conducted with \texttt{bf16} precision using a cosine annealing learning rate schedule, starting from an initial learning rate of $1 \times 10^{-5}$ for a total of 1.1 epochs, with a warm-up ratio of 10\%.

For the GRPO stage, we follow the original algorithm's recommended hyperparameter settings: KL regularization coefficient $\beta = 0.001$, clipping coefficient $\varepsilon = 0.2$, and group size $G = 8$. This configuration has been shown in prior work to reduce training cost while maintaining stability. The weighting coefficients $\lambda_{\text{format}}$, $\lambda_{\text{cap}}$, and $\lambda_{\text{ans}}$ are all set to $\frac{1}{3}$ to ensure equal contribution from each reward component.

All experiments are implemented using the PyTorch framework and run on 8$\times$96GB H20 GPUs.

\begin{figure*}[th]
\centering
\includegraphics[width=\textwidth]{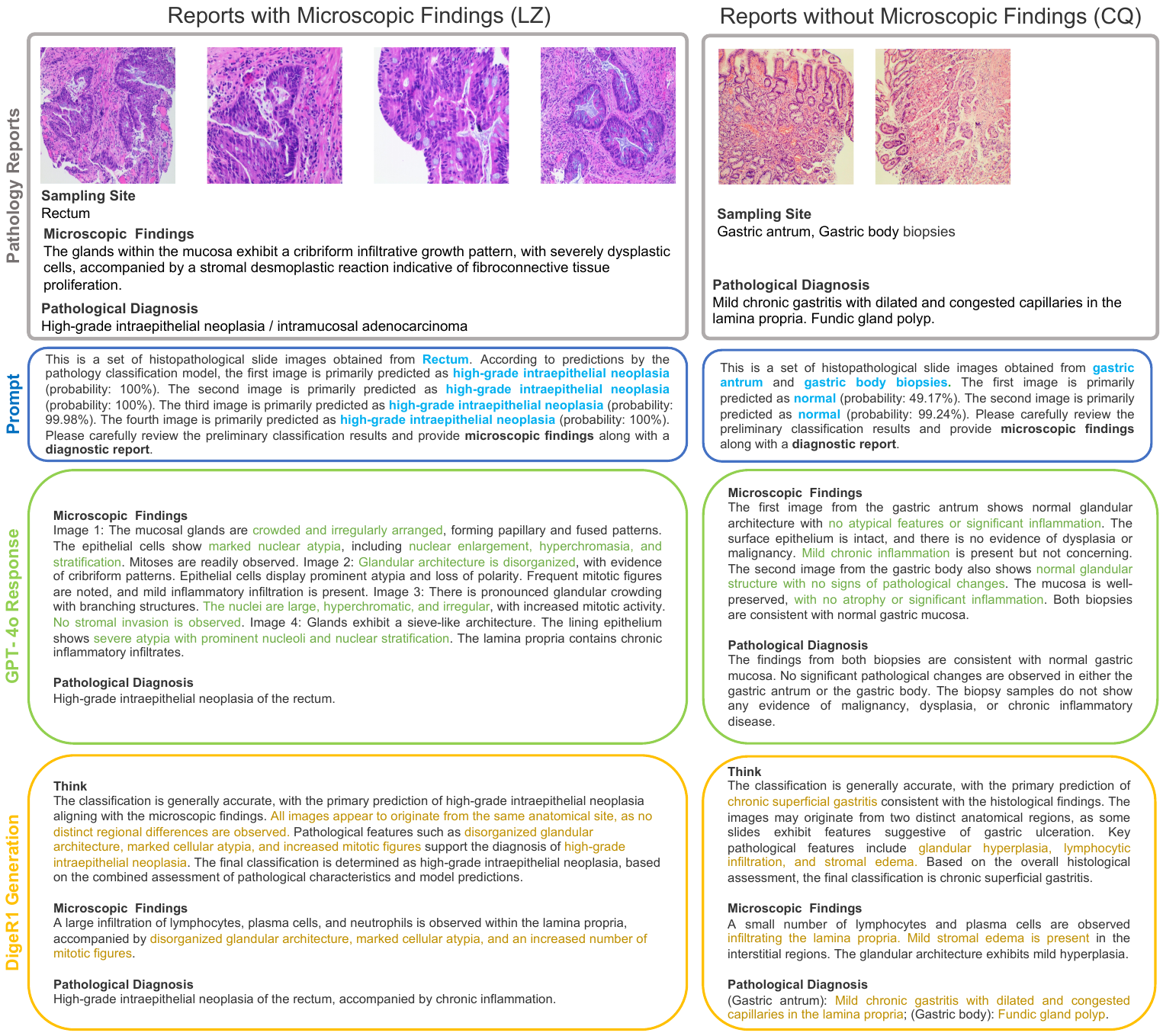}
\caption{
Comparison of generation results between DiagR1 and GPT-4o on two test datasets. After applying prompt argumentation to the reports from both datasets, the inputs were processed by the models, resulting in the GPT-4o response and DiagR1 generation.
}
\label{vis}
\end{figure*}

\subsection{Evaluation and Comparison}

Table~\ref{tab:evaluation1} reports the comparative results of various models on the generation of microscopic findings and diagnostic reports across two digestive pathology datasets. The evaluated baselines include the state-of-the-art proprietary models GPT-4o and Gemini-2.5-pro, along with leading open-source vision-language models such as LLaMA-3.2-11B-Vision-Instruct, DeepSeek-VL2. To assess the impact of fine-tuning strategies, we further include our base model Qwen2.5-VL-7B, its supervised fine-tuned variant DiagSFT, and the reinforcement learning enhanced variant DiagR1. All models are evaluated under a unified prompt argumentation setting to ensure consistency.

The results clearly demonstrate that DiagR1 outperforms all other models across all tasks. DiagSFT ranks second, while among models without post-training, GPT-4o shows the best performance. Specifically, on the microscopic findings generation task of LZ, DiagR1 outperforms GPT-4o by 5.90 percentage points in BERTScore and by 11.83 percentage points in BLEU. On the diagnostic report generation task of LZ, DiagR1 achieves a 7.73\% improvement in BERTScore and an 18.24\% increase in BLEU. Furthermore, on the diagnostic task of CQ, DiagR1 outperforms GPT-4o by 2.28\% and 1.81\% in BERTScore and BLEU, respectively.

\begin{table*}[h]
\centering
\caption{The impact of different prompt augmentation strategies on pathology report generation tasks. The best-performing methods are \textbf{bolded}}
\label{tab:evaluation2}
\begin{tabular}{lcccccc}
\toprule
\textbf{Model} & \multicolumn{2}{c}{\textbf{Microscopic Findings (LZ)}} & \multicolumn{2}{c}{\textbf{Diagnosis (LZ)}} & \multicolumn{2}{c}{\textbf{Diagnosis (CQ)}} \\
\cmidrule(lr){2-3} \cmidrule(lr){4-5} \cmidrule(lr){6-7}
& BERT Score (\%) & BLEU (\%) & BERT Score (\%) & BLEU (\%) & BERT Score (\%) & BLEU (\%) \\
\midrule
GPT-4o & 79.83 & 0.90 & 79.23 & 1.49 & 81.37 & 1.38 \\
GPT-4o + Pos & 81.80 & 0.89 & 80.32 & 1.07 & 81.09 & 1.93 \\
GPT-4o + Pos + Cls & 82.66 & 1.12 & 85.07 & 2.69 & 84.37 & 2.39 \\
\midrule
Qwen2.5-VL 7B & 75.32 & 0.50 & 72.10 & 0.50 & 73.57 & 0.60 \\
Qwen2.5-VL 7B + Pos & 78.31 & 1.21 & 73.31 & 0.76 & 82.40 & 1.55 \\
Qwen2.5-VL 7B + Pos + Cls & 81.35 & 1.12 & 79.65 & 0.90 & 81.13 & 1.53 \\
\midrule

DiagSFT & 86.38 & 7.55 & 88.39 & 7.61 & 84.73 & 3.42 \\
DiagR1 & \textbf{88.56} & \textbf{12.95} & \textbf{92.80} & \textbf{20.93} & \textbf{86.65} & \textbf{4.20} \\
\bottomrule
\end{tabular}
\end{table*}

\begin{table*}[h]
\centering
\caption{The impact of the Chain-of-Thought mechanism on pathology report generation tasks. The best-performing methods are \textbf{bolded}}
\label{tab:evaluation3}
\begin{tabular}{lcccccc}
\toprule
\textbf{Model} & \multicolumn{2}{c}{\textbf{Microscopic Findings (LZ)}} & \multicolumn{2}{c}{\textbf{Diagnosis (LZ)}} & \multicolumn{2}{c}{\textbf{Diagnosis (CQ)}} \\
\cmidrule(lr){2-3} \cmidrule(lr){4-5} \cmidrule(lr){6-7}
& BERT Score (\%) & BLEU (\%) & BERT Score (\%) & BLEU (\%) & BERT Score (\%) & BLEU (\%) \\
\midrule
Qwen2.5-VL 7B & 81.35 & 1.15 & 79.65 & 0.90 & 81.13 & 1.53 \\
\midrule
DiagSFT (w/o \texttt{<think>}) & 84.32 & 5.61 & 82.44 & 7.86 & 82.43 & 1.66 \\
DiagSFT & 86.38 & 7.55 & 88.39 & 7.61 & 84.73 & 3.42 \\
\midrule
DiagR1 (w/o \texttt{<think>}) & 87.17 & 8.51 & 84.55 & 8.06 & 83.99 & 3.56 \\
DiagR1 & \textbf{88.56} & \textbf{12.95} & \textbf{92.80} & \textbf{20.93} & \textbf{86.65} & \textbf{4.20} \\
\bottomrule
\end{tabular}
\end{table*}

\subsection{Visualization and Interpretation}

To systematically evaluate the models’ performance in real-world pathology report generation tasks, we selected two representative samples from the test sets: one from LZ, which contains expert-annotated microscopic findings (left panel), and another from CQ, which includes only the pathological diagnosis without microscopic descriptions (right panel). Prompt-enhanced inputs were applied to both reports and submitted to GPT-4o and DiagR1, with the corresponding outputs shown in Fig.~\ref{vis}.

From a detailed perspective, GPT-4o exhibits a certain degree of hallucination in generating microscopic findings. For example, it often introduces irrelevant or unsupported histological content. Moreover, its output tends to be verbose and repetitive, and the pathological diagnosis is imprecise, failing to effectively map model predictions to corresponding histological evidence, lacking both specificity and clinical usability.

In contrast, DiagR1 produces more structured, concise, and clinically aligned outputs that better adhere to the writing conventions of real pathology reports. It begins with semantic reasoning over the prompt through the \texttt{<think>...</think>} module, identifying heterogeneity in image structure and reasonably inferring that the slides may originate from different anatomical regions (e.g., distinct segments of the terminal ileum). It then explains the logic connecting predicted labels such as low-grade intraepithelial neoplasia and non-neoplastic tissue, followed by the generation of medically appropriate microscopic descriptions (e.g., “glandular hyperplasia with loss of nuclear polarity”) and precise diagnostic conclusions that are both domain-specific and fact-consistent.

Even in scenarios where the original data lack microscopic findings, DiagR1 can still generate structured histological descriptions based on image-level predictions and contextual information, demonstrating strong clinical reasoning and multimodal completion capabilities. Compared to GPT-4o, DiagR1 shows clear advantages in output accuracy, structural clarity, and diagnostic interpretability, highlighting its superior pathology-specific language generalization and deployment potential.

\subsection{Ablation study}

\subsubsection{Prompt Argumentation} 

To systematically evaluate the effectiveness of the Prompt Argumentation strategy in pathological report generation, we conducted a series of controlled experiments using three progressively enriched prompt configurations. The first setting employs a basic prompt that directly queries microscopic findings and pathological diagnoses, as illustrated in Fig.~\ref{start}. The second adds specimen location information to the prompt (denoted as “+pos” in Table 2), while the third further incorporates classification outputs from a pre-trained pathology model (“+cls”). We assessed these configurations using GPT-4o, our base open-source model Qwen2.5-VL 7B, and two fully fine-tuned models—DiagSFT and DiagR1—across multiple test sets. The results are presented in Table~\ref{tab:evaluation2}.

The findings indicate that adding structured contextual information to prompts significantly enhances model performance. In the diagnosis generation task on LZ, GPT-4o’s BERTScore improved from 79.23\% with the basic prompt to 85.07\% when both location and classification information were included, reflecting a relative gain of approximately 7.40\%. The BLEU score rose from 1.49\% to 2.69\%, an increase of over 80\% in surface-level lexical match. A similar improvement was observed for Qwen2.5-VL 7B, where the BERT Score increased from 72.10\% to 79.65\%, and BLEU from 0.50\% to 0.90\%, confirming the consistent benefit of enriched prompts across model architectures.

Among all models, DiagR1 achieved the highest scores across tasks and datasets. It yielded a BLEU score of 12.95\% for microscopic findings and 20.93\% for diagnostic statements on LZ, as well as 4.20\% on CQ. These results surpass both commercial and open-source baselines in terms of semantic alignment and fluency. These experiments demonstrate the critical role of prompt design in medical language generation. Incorporating spatial context and model-derived priors into prompts not only improves the semantic fidelity of generated text but also enhances its clinical relevance.

\subsubsection{Effect of Chain-of-Thought Prompting} 

To further investigate the impact of the Chain-of-Thought (CoT) mechanism on pathological report generation, we designed a controlled comparison between two prompt settings. In one setting, we explicitly inserted \texttt{<think>...</think>} tags into the prompt to encourage intermediate reasoning before producing the final answer. In the other, this reasoning step was omitted, allowing the model to directly generate the output. This setup enables us to isolate the contribution of explicit reasoning cues to generation quality.

We applied both supervised fine-tuning (SFT) and reinforcement learning (GRPO) to the base model Qwen2.5-VL 7B, under conditions with and without CoT guidance, yielding four model variants: DiagSFT (w/o \texttt{<think>}), DiagSFT, DiagR1 (w/o \texttt{<think>}), and DiagR1. These models were evaluated on three tasks: microscopic findings and diagnostic report generation on LZ, and diagnostic report generation on CQ. The results are summarized in Table~\ref{tab:evaluation3}.

The results clearly demonstrate the effectiveness of the CoT prompting strategy. For instance, in the diagnostic report generation task on LZ, DiagR1 with CoT achieved a BERTScore of 92.80\%, compared to 84.55\% without CoT—a relative improvement of approximately 9.80\%. BLEU scores improved even more substantially, from 8.06\% to 20.93\%. On CQ, the BERTScore rose from 83.99\% to 86.65\%, while BLEU increased from 3.56\% to 4.20\%, further confirming the positive effect of CoT prompts on generalization performance.

Even in the absence of CoT prompting, both DiagSFT and DiagR1 outperformed the base model Qwen2.5-VL 7B by a significant margin, indicating that the fine-tuning strategies themselves effectively enhanced the model's report generation capabilities. The addition of explicit reasoning chains, however, further elevated the models’ semantic coherence and logical structure.

\section{Conclusion}

This study proposes a multimodal post-training framework for gastrointestinal pathology report generation, addressing the challenges of incomplete structural information and factual hallucinations. By constructing a structured pathology dataset and incorporating anatomical site and lesion classification as contextual prompts, the model is guided to better capture image semantics and generate clinically coherent text. The training pipeline combines supervised fine-tuning with GRPO, enabling the model to produce structured, semantically consistent, and interpretable reports. Experimental results demonstrate that the proposed DiagR1 model significantly outperforms existing open-source and proprietary baselines across multiple tasks, particularly in generating microscopic findings and diagnostic conclusions.

Despite these promising results, certain limitations remain. First, the current evaluation is conducted in an offline setting, without integration into real-world clinical workflows. Second, the generalization ability of the model has not yet been fully validated across diverse disease types and data distributions.

Future work will focus on evaluating the model's robustness under cross-institutional and heterogeneous imaging conditions and conducting small-scale pilot studies in clinical settings, laying the groundwork for its practical application in assisted pathology reporting and diagnostic support.



\section*{References}

\nocite{*}
\bibliography{ref} 
\bibliographystyle{ieeetr} 

\end{document}